# Structure of optical vortices produced by holographic gratings with "fork" geometry: Kummer beams

A. Ya. Bekshaev, A. I. Karamoch

I. I. Mechnikov National University, Dvorianska 2, Odessa, 65026 Ukraine

M. V. Vasnetsov, V. A. Pas'ko, M. S. Soskin

Institute of Physics, National Academy of Sciences, Prospect Nauki 46, Kiev, 03028 Ukraine

**Abstract**

Holographic gratings with topological defects (branching of one or more fringes) are widely used for generating light beams with optical vortices (OV). This work presents an analysis of OV beams produced by binary computer-generated holograms enlightened by Gaussian beams centered at the fringe bifurcation point. Usually such beams are considered as analogs of the standard solutions of paraxial wave equation – Laguerre-Gaussian (LG) modes representing classical examples of OVs. However, the intensity profile and the whole process of their spatial evolution show important differences from the LG prototypes. In the case of integer topological charge, a created OV beam can be described by the special Kummer function, which allows referring to these beams as to "Kummer beams". Properties of Kummer beams are studied numerically and analytically. Main distinctions from the corresponding LG modes are much slower intensity decay at the beam periphery and much higher beam divergence; differences between the Kummer and LG beams grow with the OV topological charge.

Light beams with optical vortices (OV) possess many interesting properties and are the objects of intensive research activity.[1-9] Investigations in this area were initiated by basic papers of J. Nye and M. Berry, who introduced a concept of wavefront dislocations into the wave theory.[1,2] Authors chose this term due to close analogy with dislocations in crystals. An electromagnetic wave can possess phase defects along continuous lines, where the wave amplitude vanishes. As M. Berry has noted, there are three interpretations of these lines: as wavefront dislocations, since the patterns of constant-phase surfaces (wavefronts) mirror those of dislocations in the arrangements of atoms in crystals; as vortices, since the phase gradient direction (that is, direction of the energy current, or of the Poynting vector) swirls about the singular line like fluid in an irrotational vortex; and as zeros, that is "threads of darkness".[4] The spiral-like pattern of the energy flow justifies the term "optical vortex" that was introduced by P. Coullet, L. Gil and F. Rocca to describe the light field in a laser cavity,[5] and now is widely used in optical science.

One of the most important features of an OV is the helical wavefront configuration expressed by the beam phase spatial dependence in the form

$$\Phi(\varphi, z) = kz + m\varphi \tag{1}$$

where $z$ is the propagation axis, $\varphi$ is the azimuth angle in the beam cross section, $m$ is signed integer named topological charge of an OV and $k$ is the radiation wave number. In 1992, a group at Leiden University in the Netherlands recognized that light beams with helical wavefronts carry the mechanical angular momentum with respect to the propagation axis (so called orbital angular momentum) which equals $m\hbar$ per photon.[10] This property of OV beams constitutes a great interest for fundamental science and also find practical applications. The dark OV core serves a physical instrument to trap and manipulate small particles;[11] the orbital angular momentum can be transferred to trapped objects causing their rotation.[12]

A simple "classical" case of a $m$-charged OV beam is the Laguerre-Gaussian mode $LG_0^m$. Employing the usual representation of the electromagnetic field of a paraxial beam through the slowly varying complex amplitude $u(r, \varphi, z)$ where $r$ is the polar radius so that $(r, \varphi)$ forms a polar coordinate frame in the beam cross section,

$$E(r,\varphi,z) = u(r,\varphi,z)\exp(ikz),$$

we can describe the $LG_0^m$ mode by expression

$$u(r,\varphi,z) = \frac{E_0}{\sqrt{|m|!}} \frac{w_0}{w} \left(\frac{r}{w}\right)^{|m|} \exp\left(-\frac{r^2}{w^2}\right) \exp\left\{i\left[\frac{kr^2}{2R} + m\varphi - (|m|+1)\arctan\frac{z}{z_R}\right]\right\}, \qquad (2)$$

where $E_0$ is the amplitude parameter, $w = w_0\sqrt{1+z^2/z_R^2}$ is the current beam radius, $w_0$ is the beam radius at the waist supposed to coinside with the plane $z = 0$, $R = z(1 + z_R^2/z^2)$ is the wavefront curvature radius, $z_R = kw_0^2/2$ is the beam Rayleigh length.[13]

Laser beams with single vortices can be obtained directly from a laser with some modifications, but this way is the least reliable and poorly controllable, so other techniques to create such kind of beams were developed. One can produce a helical wave by means of the special phase mask with helical relief introduced into a beam with smooth wavefront.[14] More usual "rectangular" Hermite-Gaussian modes can be transformed into corresponding LG modes with the help of so called mode converters.[15,16] The most common and widespread method for creating helical beams is the use of computer-generated gratings.[17-19] The idea of OV beam formation by use of diffraction of an ordinary wave on a computer-generated grating is based on the holographic principle: a readout beam restores the wave, which participated in the hologram recording. Instead of writing a hologram with two actual optical waves, it is enough to calculate the interference pattern numerically and print the picture in black and white or grayscale. Then the picture after reduction of the transverse dimensions serves as amplitude grating producing necessary OV beam in the diffraction order.

The pattern of interference between an OV beam and a beam with regular wavefront, e.g., plane wave, with slightly disagreed directions, has remarkable difference from the usual picture of equidistant fringes. The azimuthal phase dependence (1) results in the splitting of the central fringe into $|m|$ new fringes with formation of the "fork" (bifurcation) structure (Fig. 1). Corresponding computer-generated gratings are usually referred to as gratings with embedded $m$-charged singularity.

The grating acts on an incident beam as a transparency with the inhomogeneous transmission $T(x, y)$ determined by the phase difference between signal and reference waves at a given point of the hologram $\Delta\Phi(x, y)$ (here $x, y$ are Cartesian coordinates in the grating plane). The simplest version of the grating is an amplitude binary grating for which

$$T(x,y) = \begin{cases} 0, & \cos(\Delta\Phi) \leq 0 \\ 1, & \cos(\Delta\Phi) > 0 \end{cases}, \qquad (3)$$

which neglects the fringe contrast variation. Compared to the usual "gray" gratings, the binary ones can be prepared very accurately by a rather simple procedure. Examples of binary gratings with embedded singularity are given in Fig. 1.

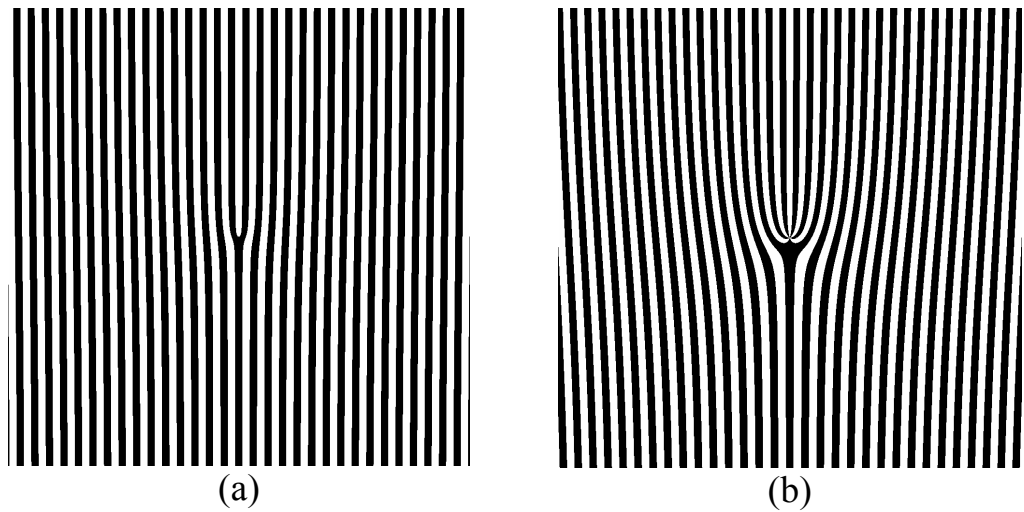

Fig. 1. The computer-generated patterns for binary gratings which are able to produce in the first diffraction order OV beams with topological charges $m = 1$ (a) and $m = 5$ (b).

Since the first computer-generated grating with characteristic "fork" was developed in 1990 by V. Bazhenov, M. Vasnetsov and M. Soskin,[17] only a few efforts have been made to theoretical description of their action. In particular, N. Heckenberg and coworkers showed that an OV beam obtained by means of diffraction of a Gaussian beam by such binary grating can be represented as a superposition of LG modes with equal topological charges but different radial indices;[19] however, this representation is not suitable because it employs infinite functional series with sometimes poor convergence. In this work, we propose the diffracted beam description in a mathematically concise and physically transparent form.

To understand properties of the light waves appearing due to diffraction of a readout Gaussian beam by the binary grating with embedded singularity, let us compare the far field behavior of the diffracted beams with that of $LG_0^m$ modes (2) with the same $m$. In both cases, the initial near-field complex amplitude distribution can be expressed in the form

$$u_m(r,\varphi) \propto u_0(r)\exp(im\varphi). \tag{4}$$

For a diffracted beam, supposing the waist of the readout Gaussian beam to coincide with the grating plane,

$$u_0(r) = \exp\left(-\frac{r^2}{w_0^2}\right), \tag{5}$$

the azimuthal phase dependence $\exp(im\varphi)$ can be imparted by a grating with $m$-charged singularity in the first diffraction order or by a grating with the unit-charge singularity in the $m$-th diffraction order.[19] For the $LG_0^m$ mode (2),

$$u_0(r) = \frac{1}{\sqrt{|m|!}}\left(\frac{r}{w_0}\right)^{|m|}\exp\left(-\frac{r^2}{w_0^2}\right). \tag{6}$$

The final beam shape in the far field is determined by the angular spectrum of the initial distribution (4). We will use the general angular spectrum definition in the form

$$U(\mathbf{p}) = \frac{k}{2\pi}\int u(\mathbf{r})\exp[-ik(\mathbf{pr})](d\mathbf{r}), \tag{7}$$

where $\mathbf{r}$ is the radius-vector in the initial beam cross section, $\mathbf{p}$ is the radius-vector in the Fourier plane. In case of circular symmetry we can introduce polar angular-spectrum coordinates $\mathbf{p} = \begin{pmatrix} p_x \\ p_y \end{pmatrix} = p\begin{pmatrix} \cos\psi \\ \sin\psi \end{pmatrix}$ so that $(\mathbf{pr}) = pr\cos(\varphi-\psi)$. Then, after substitution of Eq. (4) for $u(\mathbf{r})$, Eq. (7) gives

$$U(\mathbf{p}) = U(p,\psi) = \frac{k}{2\pi}\int_0^\infty u_0(r)r\,dr\int_0^{2\pi}\exp[im\varphi - ikpr\cos(\varphi-\psi)]d\varphi,$$

which, allowing for the known relation[20]

$$\int_0^{2\pi}\exp[i(m\varphi - kpr\cos\varphi)]d\varphi = 2\pi i^{-|m|}J_{|m|}(kpr),$$

where $J_n$ is a symbol of Bessel function, leads to representation

$$U(p,\psi) = ke^{il\psi}(-i)^{|m|}\int_0^\infty rJ_{|m|}(kpr)u_0(r)\,dr. \tag{8}$$

Then for the LG$_0^m$ mode (2), with taking into account Eq. (6), the angular spectrum is

$$U_0^m(p,\psi) = \sqrt{\frac{1}{|m|!}} z_R(-i\eta)^{|m|} e^{im\psi} \exp\left(-\frac{1}{2}\eta^2\right), \qquad (9)$$

where

$$\eta = \frac{kpw_0}{\sqrt{2}} = \frac{p}{\gamma} \qquad (10)$$

is the normalized angular frequency, $\gamma = \sqrt{2}/(kw_0)$ is the divergence angle of the Gaussian beam (5).[21]

For the diffracted beam, Eqs. (8) and (5) yield

$$U_m(p,\psi) = k e^{im\psi}(-i)^{|m|} \int_0^\infty r J_{|m|}(kpr) \exp\left(-\frac{r^2}{w_0^2}\right) dr.$$

This integral can be evaluated in the form

$$U_m(p,\psi) \equiv U_K^m(p,\psi) = \frac{\Gamma\left(\frac{|m|}{2}+1\right)}{2^{|m|/2}\Gamma(|m|+1)} z_R(-i\eta)^{|m|} e^{im\psi} M\left(\frac{|m|}{2}+1, |m|+1, -\frac{\eta^2}{2}\right), \qquad (11)$$

where $\Gamma$ is the gamma function, $M$ is the confluent hypergeometric Kummer function,[20] which allowed referring to beams produced by the Gaussian beam diffraction on a binary grating with integer topological charge as to Kummer beams.[22] This reference is emphasized in Eq. (11) by the lower index in notation $U_K^m(p,\psi)$.

However, known properties of the modified Bessel functions[23] $I_n$ enables further simplification of the Kummer beams' description:

$$U_K^m(p,\psi) = \sqrt{\frac{\pi}{8}} e^{im\psi}(-i)^{|m|} z_R \eta \exp\left(-\frac{\eta^2}{4}\right)\left[I_{\frac{|m|-1}{2}}\left(\frac{\eta^2}{4}\right) - I_{\frac{|m|+1}{2}}\left(\frac{\eta^2}{4}\right)\right]. \qquad (12)$$

In particular, when $m$ is even, result (12) can be expressed through elementary functions,[20,23] for example, $\sqrt{\pi/8}\,\eta I_{1/2}(\eta^2/4) = \sinh(\eta^2/4)$, $\sqrt{\pi/8}\,\eta I_{-1/2}(\eta^2/4) = \cosh(\eta^2/4)$, etc. Interestingly, Eq. (12) is similar to the result obtained by Berry[24] for the plane wave diffraction on a spiral phase plate and differs from it only by replacing the Bessel functions $J_n$ by the modified Bessel functions $I_n$. It is not surprising because, in accord with Eq. (4), action of the considered holographic grating is identical to that of a helical phase step, and Eq. (12) can be derived by the substitution mentioned in Ref.[24] for extension of the plane-wave results to the case of a Gaussian incident beam with allowance for the far-field conditions.

The main consequence of the obtained results (9) and (11), (12) is that in the far field both Kummer and LG$_0^m$ modes have the phase spatial dependence in the form exp($im\psi$), that is, possess perfect helical shape of the wavefront (compare to Eqs. (1), (2)). This feature unite them with the initial distributions (4) at $z = 0$ and make both families of beams appropriate representatives of the isotropic OVs.[3] Hence, the amplitude distributions following from Eqs. (9) and (12) do not depend on the polar angle $\psi$ and are functions only of the radial spatial frequency $p$:

$$\left|U_0^m(p,\psi)\right| \equiv \left|U_0^m(p)\right|, \quad \left|U_K^m(p,\psi)\right| \equiv \left|U_K^m(p)\right|.$$

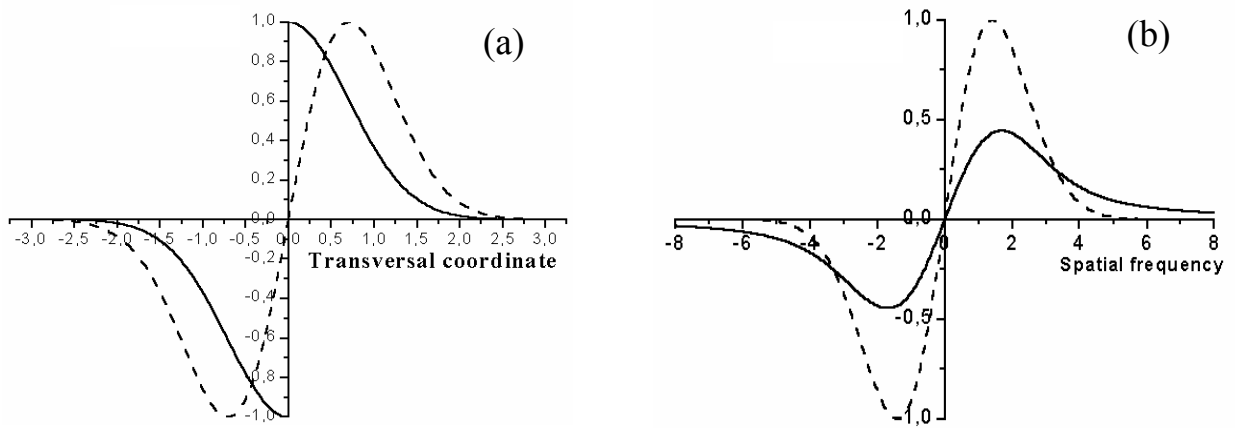

Fig. 2. Complex amplitude distributions along the axis $x$ for the Kummer beam with $m = 1$ (solid line) and for the $\mathrm{LG}_0^1$ mode (dashed line) at $z = 0$ (a) and in the far field (b). The relation of the amplitude coefficients is chosen in such a way that maximal values in initial amplitude distributions are equal.

Fig. 2 provides general comparison of a Kummer beam ($m = 1$) and a mode $\mathrm{LG}_0^1$ in the near and far field. The Kummer beam complex amplitude at $z = 0$ has a discontinuity associated with the phase jump $\pi$ upon crossing the beam axis (see Eq. (4)). This discontinuity evokes essential peculiarities in the far-field behavior that can be seen in detail in Fig. 3 presenting the normalized amplitudes of functions (9) and (12),

$$A_0^m(p) = \left|U_0^m(p)\right|/z_R, \quad A_K^m(p) = \left|U_K^m(p)\right|/z_R \qquad (13)$$

and Fig. 4 where the intensities normalized as

$$Q_0^m(p) = |m|^{1/2}\left[A_0^m(p)\right]^2, \quad Q_K^m(p) = |m|^{3/2}\left[A_K^m(p)\right]^2 \qquad (14)$$

are shown. In contrast to Fig. 2, where the compared distributions are normalized so that their maxima at $z = 0$ are equal, all curves in Figs. 3 and 4 represent beams with the same total power,

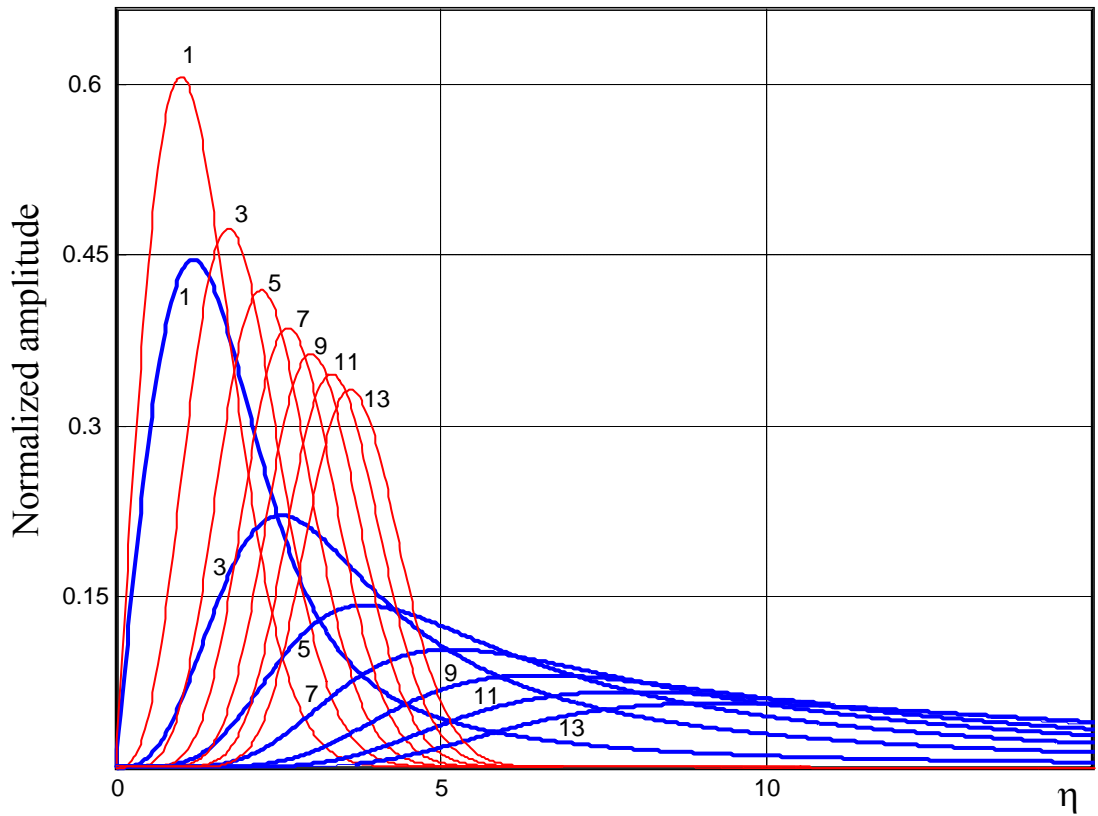

Fig. 3. Radial distributions of normalized amplitudes (13) for Kummer beams (thick lines) and for $LG_0^m$ modes (thin lines); each line is marked by corresponding topological charge $|m|$. All curves represent beams with the same total power.

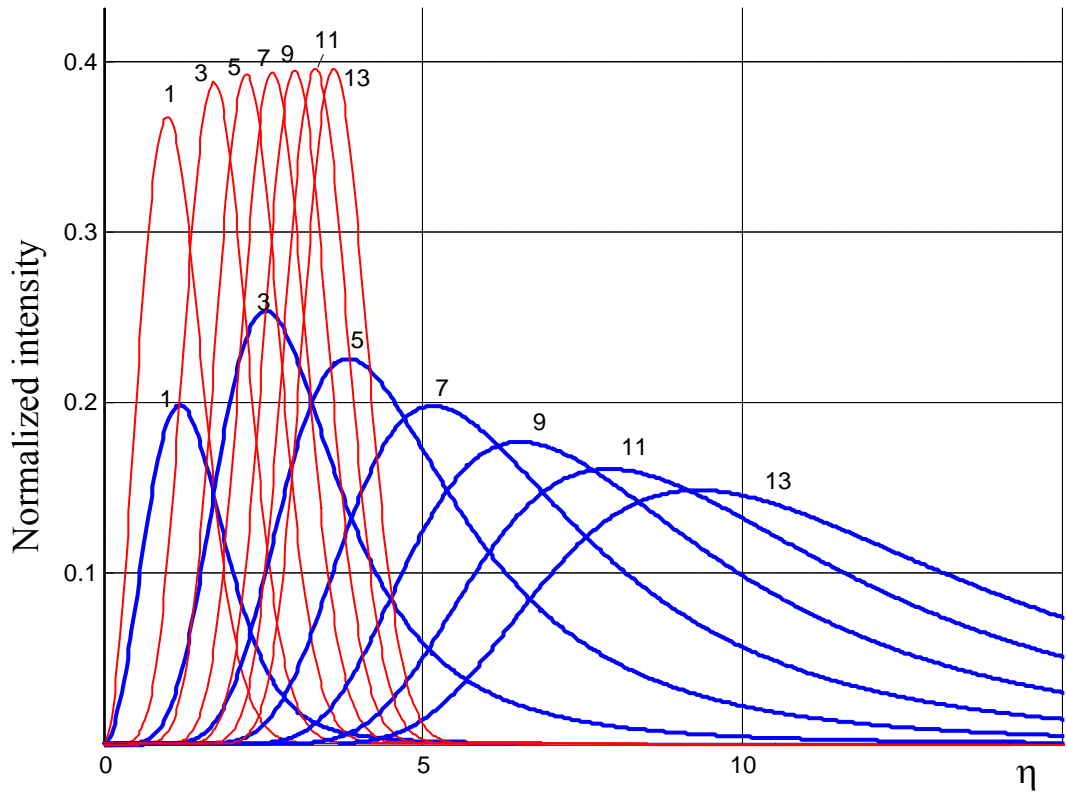

Fig. 4. Radial distributions of normalized intensities (14) for Kummer beams ($Q_K^m(p)$, thick lines) and for $LG_0^m$ modes ($Q_0^m(p)$, thin lines); each line is marked by corresponding topological charge $|m|$. All curves represent beams with the same total power.

$$\int_0^\infty \left[ A_K^m(p) \right]^2 p\, dp = \int_0^\infty \left[ A_0^m(p) \right]^2 p\, dp = \text{const}.$$

Both amplitude and intensity radial distributions of the Kummer beams differ noticeably from those of the $LG_0^m$ beams. With growing $|m|$, distinctions of the Kummer beams become more essential; they get wider and their maxima shift to higher spatial frequencies in comparison with corresponding $LG_0^m$ modes of the same topological charges. Simultaneously, the peak values of the intensity distributions decrease very rapidly. This is not seen in the intensity distributions (Fig. 4) because of the $|m|$-dependent normalization (14), which, however, makes the curves for different $m$ more comparable and enables to present their shapes clearly.

Probably, the most impressive peculiarity of the Kummer beams seen in Figs. 3, 4 is their slow fall-off at the beam periphery. In special contrast to the LG beams, all thick curves possess intensive "tails" at high η. This feature can be studied analytically by means of the asymptotic expressions[20] for $I_n$ which readily give from Eq. (12)

$$\left. U_K^m(p,\psi) \right|_{p\to\infty} = |m|(-i)^{|m|} e^{im\psi} \left( kp^2 \right)^{-1}. \tag{15}$$

Hence, instead of exponential amplitude decay typical for LG angular spectra (9), Kummer beams show much weaker transverse confinement. This is associated with higher beam divergence compared to the equally charged LG modes.

First scrutinize positions of the intensity peaks (Fig. 5). For the $LG_0^m$ modes these positions obey the well-known square-root law[3] $\eta_{\max}(m) = \sqrt{|m|}$ while for Kummer beams the linear dependence is clearly seen, which can be very well fitted by relation

$$\eta_{\max}(m) = 0.67|m| + 0.53. \tag{16}$$

Due to Eq. (10), $\eta_{\max}$ can serve a measure for the divergence of the $m$-charged Kummer beam. But the more apparent notion on the Kummer beams' divergence can be obtained from the radial coordinates at which the beam intensity falls down to a given fraction of the maximum. In Fig. 5, results for the beam radii measured at half-maximum ($\eta_{0.5}$) and 30% of maximum ($\eta_{0.3}$) levels are presented. The dependences $\eta_{0.5}(m)$ and $\eta_{0.5}(m)$ show even a slightly superlinear behavior which however can rather accurately be approximated by linear functions

$$\eta_{0.5}(m) = |m| + 1, \quad \eta_{0.3}(m) = 1.2|m| + 1.1. \tag{17}$$

The relations (16), (17) and Fig. 5 demonstrate substantial growth of the Kummer beam divergence in confrontation with $LG_0^m$ modes of the same $m$. Correspondingly, peak values of the Kummer beam decrease with growing $m$ faster than those of the $LG_0^m$ modes, which is illustrated by Fig. 6 where behavior of maxima of normalized intensities $\left[ A_K^m(p) \right]^2$ and $\left[ A_0^m(p) \right]^2$ is presented (see Eqs. (9), (12) and (13)). The peak intensity of the $LG_0^m$ mode depends on *m* in accord with the known relationship easily following from Eq. (9)

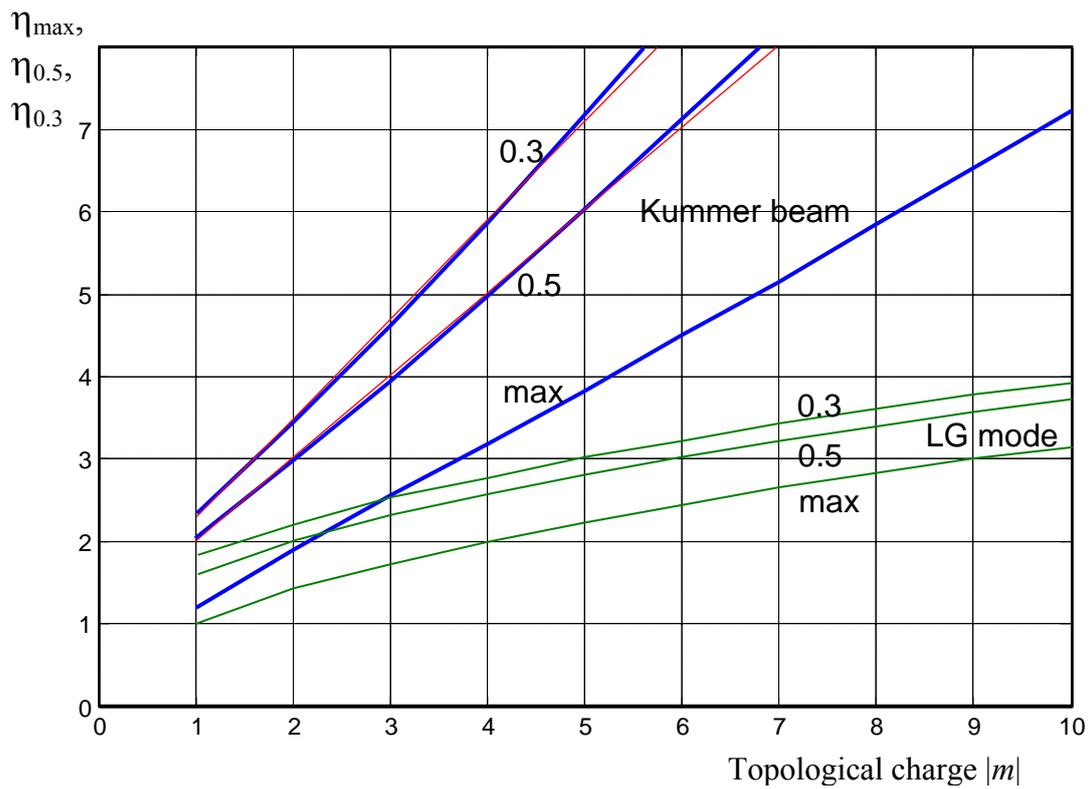

Fig. 5. Radial positions of the intensity maximums and the beam radii at levels 0.5 and 0.3 of the maximum vs the OV topological charge for Kummer beams (dark curves) and $LG_0^m$ modes (light curves) with the same initial Gaussian envelope. Thin straight lines represent approximations (17).

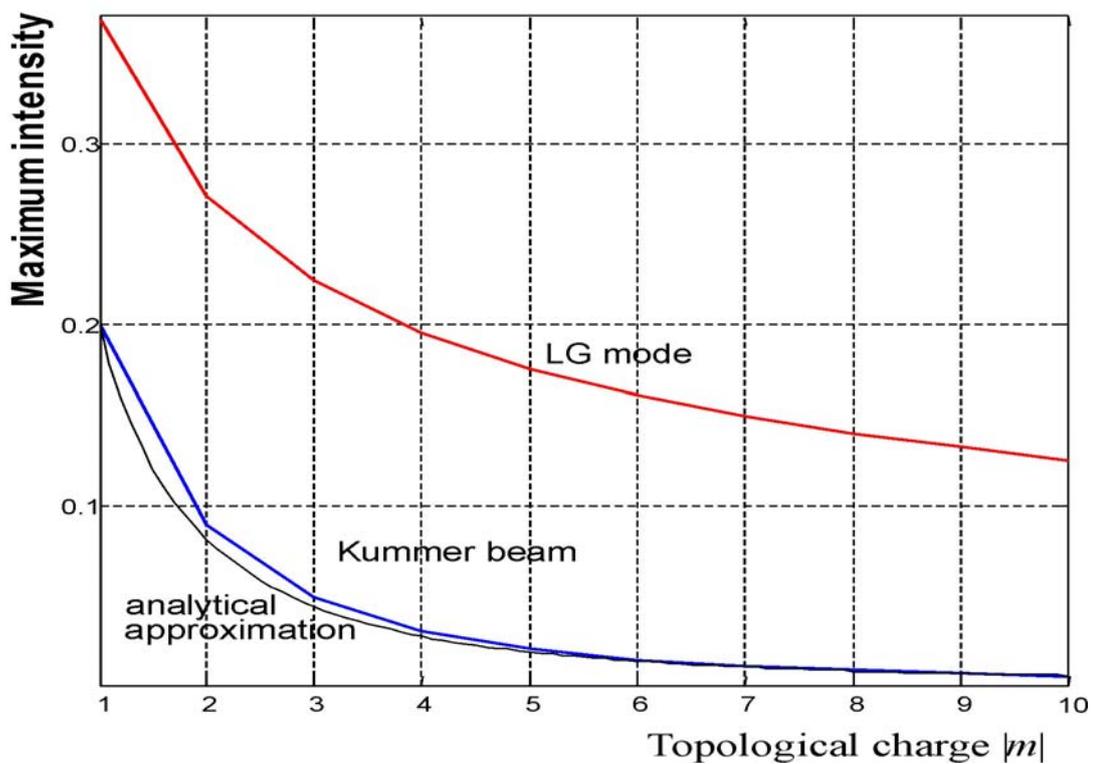

Fig. 6. Absolute values of the intensity maximum vs. the OV topological charge for Kummer beams and $LG_0^m$ modes; thin curve represents analytical approximation (19).

$$\left(\left[A_0^m(p)\right]^2\right)_{max} = \frac{1}{|m|!}|m|^{|m|}e^{-|m|}, \tag{18}$$

variation of the peak intensity of the *m*-charged Kummer beam with high accuracy can be described by equation (see Fig. 6)

$$\left(\left[A_K^m(p)\right]^2\right)_{max} = \frac{1.06}{\left(1+1.31|m|\right)^2}. \tag{19}$$

**Conclusions**

We have carried out a theoretical analysis of the optical vortex beams with integer topological charges produced by binary computer-generated holograms. It was known that such a beam can be represented as a superposition of LG modes with different radial indices. Our analysis shows that it can be described in the far field by the special mathematical function termed as confluent hypergeometric, or Kummer function. Therefore, we propose the name "Kummer beams" for such kind of beams. The properties of Kummer beams are demonstrated in comparison with corresponding LG modes. The analysis has shown that a Kummer beam has perfect helical shape of the wavefront but higher divergence compared to the LG mode with the same topological charge. The Kummer beam intensity distribution in the far field is noticeably wider and has lower maximal value than corresponding distribution of the LG beam. Moreover, OV beams obtained by holographic gratings show non-exponential transverse decay of the intensity which is expressed by significant intensity "tails" at the beam periphery. These peculiarities appear due to the grating singularity (bifurcation point) that causes discontinuity of the Kummer beam complex amplitude immediately after the hologram.

Properties of the Kummer beams are interesting from the point of view of fundamental optics and can be useful in practical applications of the OV beams produced by holographic method, for example in optical manipulation as well as in problems of information encoding and processing.[8,9,25]